\documentclass[11pt]{article}

\usepackage[english]{babel}

\usepackage[letterpaper,top=2cm,bottom=2cm,left=3cm,right=3cm,marginparwidth=1.75cm]{geometry}

\usepackage{amsmath}
\usepackage{amsthm}
\usepackage{amssymb}
\usepackage{graphicx}
\usepackage[colorlinks=true, allcolors=blue]{hyperref}

\theoremstyle{plain}
\newtheorem{thm}{Theorem}

\newtheorem{prop}[thm]{Proposition}

\theoremstyle{definition}
\newtheorem{defn}{Definition}

\theoremstyle{remark}

\title{Robustness against data loss with Algebraic Statistics}

\author{Roberto Fontana \\
Department of Mathematical Sciences \\ Politecnico di Torino \\ Corso Duca degli Abruzzi 24, 10129 Torino, Italy \\ \tt{roberto.fontana@polito.it}  \medskip \\
Fabio Rapallo \\
Department of Economics\\
  University of Genova\\
  Via Vivaldi 5, 16126 Genova, Italy \\ \tt{fabio.rapallo@unige.it}}

\date{}

\begin{document}
\maketitle

\begin{abstract}
The paper describes an algorithm that, given an initial design $\mathcal{F}_n$ of size $n$ and a linear model with $p$  parameters, provides a sequence $\mathcal{F}_n \supset \ldots \supset \mathcal{F}_{n-k} \supset \ldots \supset \mathcal{F}_p$ of nested \emph{robust} designs. The sequence is obtained by the removal, one by one, of the runs of $\mathcal{F}_n$ till a $p$-run \emph{saturated} design $\mathcal{F}_p$ is obtained. 
The potential impact of the algorithm on real applications is high. The initial fraction $\mathcal{F}_n$ can be of any type and the output sequence can be used to organize the experimental activity. The experiments can start with the runs corresponding to $\mathcal{F}_p$ and continue adding one run after the other (from $\mathcal{F}_{n-k}$ to $\mathcal{F}_{n-k+1}$) till the initial design $\mathcal{F}_n$ is obtained. In this way, if for some unexpected reasons the experimental activity must be stopped before the end when only $n-k$ runs are completed, the corresponding $\mathcal{F}_{n-k}$ has a high value of robustness for $k \in \{1, \ldots, n-p\}$.
The algorithm uses the circuit basis, a special representation of the kernel of a matrix with integer entries. The effectiveness of the algorithm is demonstrated through the use of simulations.
\end{abstract}

\section{Introduction}

Optimal designs and orthogonal fractional factorial designs are frequently used in many fields of application, including medicine, engineering and agriculture. They offer a valuable tool for dealing with problems where there are many factors involved and each run is expensive. The literature on the subject is extremely rich. A non-exhaustive list of references includes \cite{bailey2008design} for design of experiments in general,  \cite{atkinson2007optimum}, \cite{fedorov2013theory}, and \cite{pukelsheim2006optimal} for optimal designs and \cite{mukerjee2007modern}, \cite{dey2009fractional}, \cite{hedayat2012orthogonal} for orthogonal fractional factorial designs.

When searching for an optimal experimental design, we aim to select a design in order to produce the best estimates of the relevant parameters for a given sample size. There are many criteria for choosing an optimal design for the problem under study. They include alphabetical design criteria, and among these D-optimality is one of the most commonly used in applications.

In this work we focus on the notion of robustness of a design, \cite{fontana2022circuits}.  Let us suppose that a given design has $n$ runs and that the model to be estimated has $p$ parameters, with $n>p$. The model-design pair determines the design matrix $X$. The notion of robustness is important mainly for two reasons. First, the robustness of the design can be interpreted as the probability that a randomly selected subset of $p$ runs is a \emph{saturated} design (i.e the $p$ parameters of the model can be estimated). A high value of robustness has practical importance. If during the experimental activity $n-p$ runs are lost (i.e. the corresponding response values are not available) the probability that the $p$-run final design is \emph{saturated} is high. Second, thanks to the Cauchy-Binet Lemma, an high value of robustness is connected to an high value of the determinant of the information matrix $\det(X^tX)$. More specifically, if the design matrix $X$ is totally unimodular, robustness is proportional to the determinant of the information matrix and equivalent to D-optimality. 

The main result of this paper is an algorithm that starting from an initial design $\mathcal{F}_n$ (of size $n>p$) removes one by one its runs till a $p$-run \emph{saturated} design $\mathcal{F}_p$ is obtained. The choice of which point is removed at each step $k$ ($1 \leq k \leq n-p$) is aimed at finding a $(n-k)$-run sub-fraction $\mathcal{F}_{n-k}$ of the initial design with the highest value of robustness. 
The output of the algorithm is a sequence of \emph{robust} fractions $\mathcal{F}_n \supset \ldots \supset \mathcal{F}_{n-k} \supset \ldots \supset \mathcal{F}_p$. In practice, the experimental activity can start with the runs corresponding to $\mathcal{F}_p$ and continue adding one run after the other (from $\mathcal{F}_{n-k}$ to $\mathcal{F}_{n-k+1}$) till the initial design $\mathcal{F}_n$ is obtained. In this way, if for some unexpected reasons the experimental activity must be stopped before the end when only $n-k$ runs are completed,  the corresponding $\mathcal{F}_{n-k}$ has a high value of robustness, $1\leq k \leq n-p$. It is worth noting that the value of robustness of $\mathcal{F}_{n-k}$ is high \emph{for each $k$, $1\leq k \leq n-p$} as shown in the simulation study. In the simulation study, some examples are illustrated to prove the effectiveness of the algorithm. The problem of partial availability of data and techniques to prevent possible loss of information are addressed in, e.g., \cite{butler07}, \cite{dey:93}, \cite{street}, both in model-based and model-free frameworks. In the paper \cite{fontana|rapallo:19} a combinatorial approach is introduced for the analysis of orthogonal arrays with removed runs using aberrations and the Generalized Word-Length Pattern criterion.

The algorithm introduced in this paper uses the circuit basis, i.e., a special representation of the kernel of a matrix with integer entries. Exploiting some combinatorial properties of such a basis, we obtain a sequence of nested designs with high performance in terms of robustness with a unique computation of the circuit basis in the first step. The theory of robustness based on circuits is fully described in \cite{fontana2022circuits}, while the estimability of saturated designs using circuits is studied in \cite{fontana|etal:14}. 

There are no restrictions on the way in which the initial design $\mathcal{F}_n$ is determined. It can be an orthogonal fractional factorial design, a D-optimal design, or any other design defined according to the user's preferences. The only restriction applies to the design matrix $X$ that must have integer entries. It follows that ANOVA-type models for qualitative variables can be considered. Polynomial models for continuous variables can also be considered with the restriction that the entries of the design are rational numbers. We will see with examples that, from the combinatorial point of view, in some cases quantitative factors are easily rewritten in the qualitative framework. The general case of quantitative factors where an approximation of the design matrix is needed falls outside the scope of the present paper, and we will provide some insight in the concluding remarks.

The paper is organized as follows. In Section \ref{sect:circ-rob} the definition of circuit basis is introduced together with its main algebraic and combinatorial properties, and the connections with robustness of a design are reviewed. Section \ref{sect:algo} is devoted to the description of the proposed algorithm and some computational remarks. A first example on a small design is illustrated with full details. In Section \ref{sect:ex} some examples are presented and discussed. Finally, Section \ref{sect:fin} contains some final comments and pointers to future works.

\section{Circuits and robustness} \label{sect:circ-rob}

We consider a design ${\mathcal F}$ with $n$ runs, chosen from a set ${\mathcal D}$ with $N$ runs, $N>n$. For experiments with $d$ discrete factors $X_1, \ldots, X_d$, the set ${\mathcal D}$ is usually represented as a Cartesian product such as
\[
\{0, \ldots, s_1-1 \} \times \cdots \times \{0, \ldots , s_d-1\}
\]
where $s_1, \ldots, s_d$ are the number of levels of the factors $X_1, \ldots, X_d$, respectively. However, for our theory the special coding of the factor levels, and even the Cartesian product structure of the full-factorial design are irrelevant, and we may simply assume that from a large set labeled $\{1, \ldots, N\}$ a subset of $n$ runs has been selected. In the language of fractional factorial designs, the design ${\mathcal F}$ is a fraction, while the large set ${\mathcal D}$ is a full-factorial design. 

Given a full-factorial design ${\mathcal D}$ we consider a linear model on ${\mathcal D}$:
\begin{equation}\label{mod:full}
{\bf y} = X_{\mathcal D}\boldsymbol{\beta} + \boldsymbol{\varepsilon} \, ,
\end{equation}
where ${\bf y}$ is the vector containing the response variable, $X_{\mathcal{D}}$ is the model matrix, $\boldsymbol{\beta}$ is the vector of parameters, and $\boldsymbol{\varepsilon}$ is the error term. Without loss of generality, to simplify some algebraic issues of our theory, we assume that the matrix $X_{\mathcal D}$ is full-rank with dimension $N \times p$, where $p$ is the number of estimable parameters.

When a fraction ${\mathcal F}$ is selected, the expression of the model in Eq.\eqref{mod:full} becomes
\begin{equation}\label{mod:red}
{\bf y} = X_{\mathcal F}\boldsymbol{\beta} + \boldsymbol{\varepsilon} \, ,
\end{equation}
where the model matrix $X_{\mathcal F}$ has dimension $n \times p$ and is obtained from $X_{\mathcal D}$ by selecting only the rows pertaining the chosen runs.

As pointed out in the Introduction, we aim at defining an algorithm which gives ``optimal'' subsets of a given fraction. In this paper we use the criterion of robustness. Following \cite{ghosh:79} and  \cite{ghosh:82}, the robustness is defined in terms of saturated fractions.

\begin{defn}
Let ${\mathcal F}$ be a fraction with model matrix $X_{\mathcal F}$. The robustness of the fraction ${\mathcal F}$ under the model $X_{\mathcal F}$ is the proportion of saturated minimal fraction over the number of minimal fractions:
\begin{equation}
    r(X_{\mathcal F}) = \frac {\# \{\mathrm{saturated} \ {\mathcal F}_p\}}{\binom{n}{p}}
\end{equation}
\end{defn}

We observe that the number of runs of a minimal fraction is $p$ and for a minimal fraction $\mathcal{F}$ the robustness can be either $0$ or $1$.

We now show how to use the circuits of the model matrix $X_{\mathcal F}$ to study the robustness of the fraction. Then, in the next section we will provide an algorithm to sequentially remove runs from ${\mathcal F}$ to maintain the robustness as high as possible.

To match the language of Combinatorics with the design theory, we work with the transposed of the model matrix, i.e., we consider the matrix $A_{\mathcal F} = X^t_{\mathcal F}$ and with a slight abuse of notation we still call it the model matrix. Note that working with $A_{\mathcal F}$ implies that the runs identify columns, while parameters identify rows.

In words, a circuit of $A_{\mathcal F}$ is an element ${\bf u}$ of $\ker(A_{\mathcal F})$ with integer entries and minimal support, where the support of a vector ${\bf u}$ is the set of indices $i$ with $u_i \ne 0$. We denote by $\mathrm{supp}(\bf u)$ the support of the vector ${\bf u}$.

\begin{defn} \label{def:circ}
Let ${\bf u} \in {\mathbb{Z}}^n$ be an $n$-dimensional integer vector. ${\bf u} \in \ker(A_{\mathcal F})$ is a circuit if the nonzero entries of ${\bf u}$ are relatively prime and there is no other vector $v \in \ker(A_{\mathcal F})$ with $\mathrm{supp}(v) \subset \mathrm{supp}(u)$.
\end{defn}

The set of all the circuits of $A_{\mathcal F}$ is called the circuit basis of $A_{\mathcal F}$ and is denoted with ${\mathcal C}(A_\mathcal{F})$. The circuit basis is always finite. 

For a comprehensive introduction to circuits and its properties the reader can refer to \cite{ohsugi|hibi:13} and \cite{sturmfels:96}. Here, the key issue for using the circuit basis as a special basis of $\ker(A_\mathcal{F})$ is given by the following property.

\begin{prop}\label{prop:subset}
Let ${\mathcal F}'$ be a sub-fraction of ${\mathcal F}$, and decompose each circuit of $A_{\mathcal F}$ into ${\bf u}=({\bf u}_{\mathcal{F}'},{\bf u}_{\mathcal{F}-\mathcal{F}'})$. The circuits of $A_{\mathcal{F}'}$ are
\[
\{ {\bf u}_{\mathcal{F}'} \ : \ {\bf u} \in {\mathcal C}(A_{\mathcal F}), \mathrm{supp}({\bf u}) \subseteq \mathcal{F}' \} \, .
\]
\end{prop}

Prop.~\ref{prop:subset} says that the circuit basis is the natural representation of the kernel $\ker(A_{\mathcal F})$ when we need to remove runs, because we don't need to recompute the basis of the kernel at each step. The first computation contains all the information needed to compute the robustness also for all possible sub-fractions.

Moreover, the minimality property established in Def.~\ref{def:circ} is used in the following result, from \cite{fontana|etal:14}, which characterize the saturated minimal fractions using the circuit basis.

\begin{prop} \label{prop:FRR}
Let $A_{\mathcal F}$ be a model matrix with circuit basis $\mathcal{C}(A_{\mathcal F})$. A minimal sub-fraction ${\mathcal F}_p$, i.e., a fraction with $p$ runs from $\mathcal{F}$, is saturated if and only if it does not contain any of the supports $\mathrm{supp}({\bf u})$ for all ${\bf u} \in \mathcal{C}(A_{\mathcal F})$.

\end{prop}

Exploiting Prop.~\ref{prop:FRR}, in \cite{fontana2022circuits} an algorithm for finding robust fractions using the design points of a candidate set $\mathcal{D}$, based on an exchange-type strategy is described. Without introducing all the technical details, the algorithm in \cite{fontana2022circuits} is based on two rules: (i) remove the point contained in the largest number of circuits, (ii) remove the point contained in the smallest circuits. Such two rules are summarized in the definition of a loss function $L(P)$ defined as
\begin{equation}\label{eq:app}
L(P) = \sum_{{\mathbf u}} \binom{n-\#\mathrm{supp}({\mathbf u})}{p-\#\mathrm{supp}({\mathbf u})}
\end{equation}
where the sum is taken over all the circuits $({\mathbf u})$ in the fraction containing the point $P$. As noticed in \cite{fontana2022circuits}, the formula in Eq.~\eqref{eq:app} is a first-order approximation of the inclusion-exclusion formula. 

The algorithm is made by the following main steps: (a) Start from a fraction $\mathcal{F}$; (b) Remove from $\mathcal{F}$ the run with the highest loss function. (c) Add a new run from ${\mathcal D}$ not in ${\mathcal F}$. (d) Repeat steps (b)-(c) until no reduction in the number of circuits is possible. 

To actually compute the circuit basis ${\mathcal C}(A_{\mathcal F})$ for a given model matrix $A_{\mathcal F}$ there are several available packages and free software. For the computations in this paper we have used {\tt 4ti2}, see \cite{4ti2}, a program for computing combinatorial objects like Markov bases, Graver bases, circuits, and more. It is available as a stand-alone executable program, or as a package inside {\tt Macaulay2}, see \cite{M2}, a free software for Computational Commutative Algebra.

A common drawback of Algebraic Statistics tools is the limitation to small problems. The computation of the circuits does not make exception. The number of circuits for a model matrix on the full-factorial design increases fast with the number of runs and the computations are actually feasible only for small-sized problems. To give a rough idea, the circuits for a full-factorial $2^d$ design with main effects and first-order interactions is feasible only for $d \le 6$. For the $2^d$ with only main effects, the circuit basis has 20 elements for $d=3$; 1,348 elements for $d=4$; 353,616 elements for $d=5$.

As a running example, let us consider a $3 \cdot 2^2$ model with main effects and the interaction between the two binary factors. The full-factorial design has $N=12$ points and there are $p=6$ free parameters. To run the exchange-type algorithm described above, we compute the circuits of the model matrix on the full-factorial design and we obtain a circuit basis with $42$ circuits: $18$ circuits with support on $4$ points and $24$ circuits with support on $6$ points. Now, to find a robust fraction with a fixed size, it is enough to run the algorithm above with an arbitrary starting fraction with $8$ runs. One can start with a randomly selected fraction, or can select a fraction satisfying some given criteria, such as D-optimality. 

However, the algorithm introduced in this work does not need the computation of the circuit basis for the full-factorial design, but only the circuit basis for the starting fraction, and therefore it can be applied also in relatively large examples.

We make explicit here some computational remarks, and we will come back on these issue later after the introduction of the new algorithm in Section \ref{sect:algo}. First, the procedure above can be viewed as a first-order approximation of the inclusion-exclusion formula, and thus it is not assured that for each sample size it returns a fraction with the highest possible robustness. In the simulations described in \cite{fontana2022circuits}, for sample sizes near to minimal the performance are quite good, but for large fractions, where the number of circuits contained in more than one minimal fraction is not negligible, the algorithm may yield a fraction with low robustness. Second, the algorithm assumes that the circuit basis for the candidate set (in general the full-factorial design) is available. The computation of the circuit basis for large designs can be actually unfeasible and thus the practical applicability is usually limited to small cases. Third, the algorithm above works for a fixed sample size, and it yields non-nested fractions when applied with different sample sizes. This is due to the random addition of a new run, and to the presence of ties, i.e., runs with the same loss function to be removed.

Finally, we conclude with a key remark about the number of circuits in the circuit basis. As noticed is the examples above, the number of circuits can be rather large. But not all the circuits are necessary for finding robust designs. In fact it is known, see \cite{fontana2022circuits} for details and further references, that for a full-rank model matrix $A_{\mathcal F}$ with dimension $p \times n$ the supports of the circuits have at most $p+1$ design points. But, according to Proposition \ref{prop:FRR}, the circuits with support on $p+1$ points are not involved in the computation of the robustness, because they are not contained in any minimal fraction. Thus, it is possible to run the algorithm with a reduced circuit basis containing only the circuits with support on $p$ design points or less. In the remaining part of this paper we will use systematically such reduced circuit basis, and to avoid new notation we will use again the symbol ${\mathcal C}(A_{\mathcal F})$.

\section{Removing runs from a fraction} \label{sect:algo}

In this section we consider another version of the circuit-based algorithm for removing runs from a given design. While the procedure described in the previous section was essentially based on the first property of the circuits, summarized in Prop.~\ref{prop:FRR}, the algorithm below fully exploits the second property of the circuits, described in Prop.~\ref{prop:subset}. 

Given a fraction $\mathcal{F}$ and a model with (transposed) design matrix $A_{\mathcal{F}}$, we use the circuit basis ${\mathcal C}(A_{\mathcal{F}})$ as a ``geometric tool'' to choose the order of the points to be removed from $\mathcal{F}$ with the goal of obtaining the best possible robustness of the sub-fractions. To avoid computational problems we assume that the design matrix $A_{\mathcal{F}}$ is full-rank. 

The algorithm works as follows:
\begin{enumerate}
    \item Start from a fraction $\mathcal{F}$ with $n$ runs, and compute the circuit basis ${\mathcal C}(A_{\mathcal{F}})$;
    
    \item Compute the loss function for the runs in $\mathcal{F}$ based on ${\mathcal C}(A_{\mathcal{F}})$;
    
    \item Remove from $\mathcal{F}$ the run with the highest loss function. In case of ties, randomize among the runs with the highest loss function, and define a sub-fraction $\mathcal{F}'$ with $n-1$ runs;
    
    \item Iterate items $2$ and $3$ until the desired number of runs has been removed.
\end{enumerate}

The validity of the algorithm rests on Prop.~\ref{prop:subset}. Indeed, the circuit basis ${\mathcal C}(A_{\mathcal{F}})$ computed for the fraction $\mathcal{F}$ is valid also for all the sub-fractions $\mathcal{F}'\subset \mathcal{F}$.

Ideally the algorithm can be iterated until a saturated fraction is reached. Remember that in this case the robustness of the fraction can be either $0$ or $1$. Although robustness is useful for small designs with run size near to the minimum $p$, the algorithm works for all run sizes from $n$ to $p$. Even in the intermediate cases, where the complete enumeration of all the sub-fractions may be computationally difficult, the proposed algorithm is able to easily find robust sub-fractions because it works on the runs and not on the sub-fractions. 

Let us illustrate now a very small example in order to show the applicability of the algorithm above. We consider again the $3 \cdot 2^2$ case with main effects and the interaction between the two binary factors. The problem is to find robust sub-fractions of a D-optimal design. We use here a lexicographic order of the factor levels. In this example, we start with the fraction ${\mathcal F}_9$ which has $n=9$ runs:
\begin{align} \label{F9}
{\mathcal F}_9 = \{ & (-1,-1,+1),(-1,+1,-1),(-1,+1,+1),(0,-1,-1),\\
                    &  (0,+1,-1),(0,+1,+1), (+1,-1,-1),(+1,-1,+1),(+1,+1,-1)\} \,
\end{align}
with robustness $r(X_{{\mathcal F}_9})=0.5952$. The model has $p=6$ parameters, so we seek for sub-fractions with $8$, $7$ and $6$ runs.

There are $7$ circuits with support contained in ${\mathcal F}_9$: $3$ circuits with support on $4$ points, $4$ circuits with support on $6$ points. We point out that the relevant circuit basis now has only $7$ circuits, while the circuits basis for the full-factorial design has $42$ elements, as described in the previous section. The $7$ circuits are listed below (where the columns are ordered according to the list in Eq.~\eqref{F9}:
\begin{verbatim}
     0     0     0     1     -1     0    -1     0     1
     0     1    -1    -1      0     1     1     0    -1
     0     1    -1     0     -1     1     0     0     0
     1    -1     0    -1      1     0     1    -1     0
     1    -1     0     0      0     0     0    -1     1
     1     0    -1    -1      0     1     1    -1     0
     1     0    -1     0     -1     1     0    -1     1
\end{verbatim}

In ${\mathcal F}_9$ the highest loss function $L(R)$ is reached by the $3$ runs
\[
(-1,+1,-1),(0,+1,-1),(+1,+1,-1)
\]
We randomly choose the first run from the above list, thus defining the robust sub-fraction
\begin{align} \label{F8}
{\mathcal F}_8 = \{ & (-1,-1,+1),(-1,+1,+1),(0,-1,-1),(0,+1,-1),\\ 
                   & (0,+1,+1),(+1,-1,-1),(+1,-1,+1),(+1,+1,-1)\} \,
\end{align}
with robustness $r(X_{{\mathcal F}_8})=0.7143$.

Among the $7$ circuits above, only 3 of them still survive (the columns are ordered according to the list in Eq.~\eqref{F8}):
\begin{verbatim}
     0     0     1    -1     0    -1     0     1 
     1    -1    -1     0     1     1    -1     0 
     1    -1     0    -1     1     0    -1     1 
\end{verbatim}
To further reduce the sample size of the fraction we compute again the loss function of the $8$ remaining points. The highest value is reached by the $4$ runs:
\[
(0,-1,-1),(0,+1,-1),(+1,-1,-1),(+1,+1,-1)
\]
We randomly choose the third run in this list, and we get
\begin{align} \label{F7}
{\mathcal F}_7 = \{ & (-1,-1,+1),(-1,+1,+1),(0,-1,-1),(0,+1,-1), \\
                    & (0,+1,+1),(+1,-1,+1),(+1,+1,-1)\} \,
\end{align}
with robustness $r(X_{{\mathcal F}_7})=0.8571$.

Only $1$ circuit has support contained in ${\mathcal F}_7$ (the columns are ordered according the the list in Eq.~\eqref{F7}):
\begin{verbatim}
     1    -1     0    -1     1     -1     1 
\end{verbatim}
In the last step to reach the minimal fraction, we remove one of the $6$ runs in the support of the last circuit. We randomly choose the last run and we obtain
\[
{\mathcal F}_6 = \{(-1,-1,+1),(-1,+1,+1),(0,-1,-1),
(0,+1,-1),(0,+1,+1),(+1,-1,+1)\} \, .
\]
Of course this last fraction has robustness $1$, because it is a minimal fraction and does not contain any support of the circuits.

In this example we can compare the robustness of the sub-fractions ${\mathcal F}_8$, ${\mathcal F}_7$, ${\mathcal F}_6$ with the robustness of all the possible sub-fractions with $8,7,6$ runs respectively. There are $9$ sub-fractions of ${\mathcal F}_9$ with 8 runs: $6$ of them have robustness $0.5357$; $3$ of them have robustness $0.7143$. There are $36$ sub-fractions of ${\mathcal F}_9$ with 7 runs: $3$ of them have robustness $0$ and they are actually non-estimable; $24$ of them have robustness $0.5714$; $9$ of them have robustness $0.8571$. Finally, there are $84$ sub-fractions of ${\mathcal F}_9$ with 6 runs: $34$ of them have robustness $0$; $50$ of them have robustness $1$. We observe that our procedure has identified the highest robustness in all the steps.

Before the illustration of several examples in the next section, some computational remarks are needed to clarify the special features of the circuits for finding robust designs using our algorithm. First, the sub-fraction property of the circuits in Prop.~\ref{prop:subset} operates here in two ways. On one hand, we don't need the computation of the circuits for the full-factorial design, but only for the starting design. This makes feasible the computations also in cases where the circuits for the full factorial design cannot be computed. Moreover, the circuit basis in the first step is still valid throughout the whole algorithm, and no further computations are needed. These features allow us to use the algorithm also in intermediate-sized examples: the computation of the circuits for all the examples discussed in the next section are carried out in less than 1 second on a standard PC. Some limitations on the size of the problem still remains, as usual for most of the combinatorial algorithms in Design of Experiments: the use of approximate methods for large designs is outside the scope of this paper, as it requires new results also from the mathematical side. We will come back briefly on this issue later in Section \ref{sect:fin}.

\section{Examples} \label{sect:ex}

We consider four examples. The examples have been chosen to show the possibility to use the proposed approach in different contexts. In each example we consider a starting design with $n > p$ runs. Then we analyze the robustness of its sub-fractions which are obtained removing $1,\ldots,n-p$ points by the starting design. It follows that by removing $k$ points ($1 \leq k \leq n-p$), we must consider $\binom{n}{k}$ sub-fractions of the starting design and, for computing the robustness of each sub-fraction, we must consider $\binom{k}{p}$ size-$p$ fractions which are contained in it. In this way we obtain the exact distribution of the robustness of all the sub-fractions of size $n-k$ of the starting design. This distribution allows us to evaluate the goodness of the solutions proposed by the algorithm. When the number  $\binom{n}{k}\binom{k}{p}$ becomes too large we build an approximation of the distribution of the robustness by sampling. More specifically we consider $\min(\binom{n}{k},1000)$ sub-fractions and we evaluate the robustness of each sub-fraction by classifiyng $\min(\binom{k}{p},1000)$ fractions of size $p$ which are contained in it as saturated or not.

\subsection{Example 1: a Plackett-Burman design} \label{ex:pb}
\begin{figure}
\centering
\includegraphics[width=0.7\textwidth]{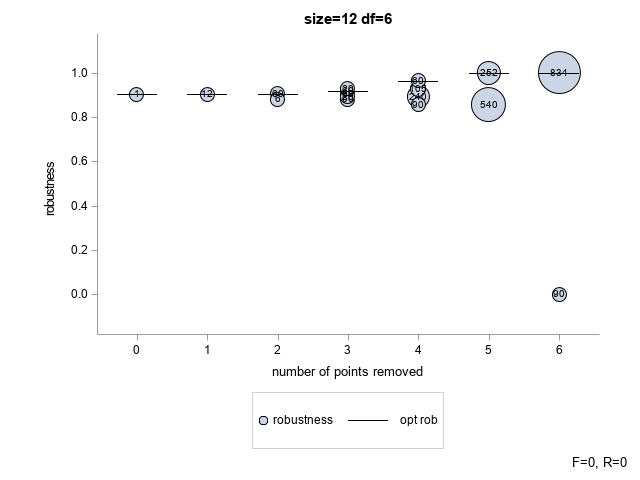}
\caption{\label{fig:bub1} Distribution of robustness \emph{vs} number of points removed for the Plackett-Burman design in Section \ref{ex:pb}. For each sample size the robustness of the design selected by the proposed algorithm is represented with an horizontal line.}
\end{figure}

The first example considers five $2$-level factors and a model with a constant term plus the $5$ main effects. The number of degrees of freedom of the model is $p=1+5=6$.  The robustness of a Plackett–Burman design with $n=12$ runs is analysed. For this problem the (reduced) circuit basis has $91$ circuits, while the corresponding (reduced) circuit basis for the full-factorial problem would consist of $44,560$ circuits. The values of the robustness of the fractions which are obtained removing $k=1,\ldots, n-p=6$ points are computed and compared with the values of the robustness corresponding to the fractions found by the algorithm. The distributions of the robustness for each $k=1,\ldots,6$ are exact. The case $k=0$ (i.e. no points removed) provides the robustness of the initial design.  The results are summarized in Figure \ref{fig:bub1}. 

Table \ref{tab:id7} compares the values of the robustness of the fractions found by the algorithm ($r_*$) with the 75th, 90th and 95th percentile of the distributions of the robustness ($p_{75},p_{90},p_{95}$ respectively). It is worth noting that for each number $k$ of points removed the algorithm provides values of robustness greater than the 75th percentile and apart from $k=3$ equal to the 95th percentile.

\begin{table}[h]
    \centering
    \begin{tabular}{c|rrrr}
    $k$ & $p_{75}$ & $p_{90}$ & $p_{95}$ & $r_*$\\
    \hline
0 & \multicolumn{4}{c}{$r_0$=0.903} \\
1 &	0.903 &	0.903 &	0.903 &	0.903 \\
2 &	0.905 &	0.905 &	0.905 &	0.905 \\
3 &	0.917 &	0.917 &	0.929 &	0.917 \\
4 &	0.929 &	0.964 &	0.964 &	0.964 \\
5 &	1 &	1 &	1 &	1 \\
6 &	1 &	1 &	1 &	1
    \end{tabular}
    \caption{Example 1: Comparison of the output of the algorithm ($r_*$) with the 75th, 90th, and 95th percentile of the distribution of the robustness ($p_{75},p_{90},p_{95}$ respectively). The value $k$ is the number of points removed by the initial design. The value corresponding to the robustness of the initial design ($r_0$) is given at $k=0$.}
    \label{tab:id7}
\end{table}

\subsection{Example 2: a $3^4$ design}\label{ex:34}
 
\begin{figure}
\centering
\includegraphics[width=0.7\textwidth]{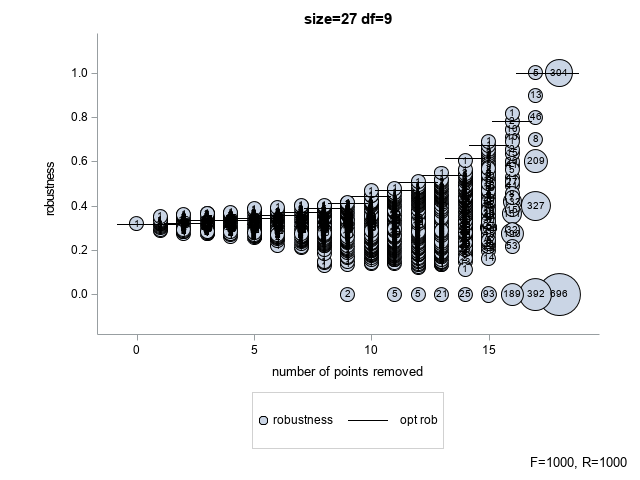}
\caption{\label{fig:bub2} Sampling distribution of robustness \emph{vs} number of points removed for the $3^4$ design in Section \ref{ex:34}. For each sample size the robustness of the design selected by the proposed algorithm is represented with an horizontal line.}
\end{figure}

The second example considers four $3$-level factors and a model with a constant term plus the $4$ main effects. The number of degrees of freedom of the model is $p=1+2\cdot4=9$.  The robustness of an orthogonal fractional factorial design with $n=27$ runs is analysed. For this problem the (reduced) circuit basis has $22,068$ circuits. The values of the robustness of the fractions which are obtained removing $k=1,\ldots, n-p=18$ points are computed and compared with the values of the robustness corresponding to the fractions found by the algorithm. The distributions of the robustness for each $k=1,\ldots,18$ are obtained by sampling. The results are summarized in Figure \ref{fig:bub2}.

Table \ref{tab:id9} compares the values of the robustness of the fractions found by the algorithm ($r_*$) with the 75th, 90th and 95th percentile of the distributions of the robustness ($p_{75},p_{90},p_{95}$ respectively). It is worth noting that for each number $k$ of points removed, apart from $k=1$, the algorithm provides values of robustness greater than the 75th percentile and for $k\geq6$ greater than the 95th percentile.

\begin{table}[h]
    \centering
    \begin{tabular}{c|rrrr}
    $k$ & $p_{75}$ & $p_{90}$ & $p_{95}$ & $r_*$\\
    \hline
0 & \multicolumn{4}{c}{$r_0$=0.308} \\
1 &	0.322 &	0.336 &	0.337 &	0.319 \\
2 &	0.329 &	0.338 &	0.345 &	0.324 \\
3 &	0.33 &	0.34 &	0.346 &	0.334 \\
4 &	0.331 &	0.34 &	0.348 &	0.336 \\
5 &	0.331 &	0.343 &	0.35 &	0.346 \\
6 &	0.334 &	0.348 &	0.355 &	0.356 \\
7 &	0.338 &	0.353 &	0.361 &	0.37 \\
8 &	0.341 &	0.358 &	0.368 &	0.392 \\
9 &	0.344 &	0.363 &	0.374 &	0.415 \\
10 &	0.349 &	0.375 &	0.389 &	0.443 \\
11 &	0.359 &	0.387 &	0.407 &	0.469 \\
12 &	0.369 &	0.406 &	0.422 &	0.509 \\
13 &	0.376 &	0.418 &	0.448 &	0.537 \\
14 &	0.4 &	0.455 &	0.492 &	0.614 \\
15 &	0.409 &	0.491 &	0.55 &	0.673 \\
16 &	0.418 &	0.582 &	0.6 &	0.782 \\
17 &	0.6 &	0.6 &	0.8 &	1 \\
18 &	1 &	1 &	1 &	1 
    \end{tabular}
    \caption{Example 2: Comparison of the output of the algorithm ($r_*$) with the 75th, 90th, and 95th percentile of the distribution of the robustness ($p_{75},p_{90},p_{95}$ respectively). The value $k$ is the number of points removed by the initial design. The value corresponding to the robustness of the initial design ($r_0$) is given at $k=0$.}
    \label{tab:id9}
\end{table}

\subsection{Example 3: a design with quantitative continuous factors} \label{ex:five}
\begin{figure}
\centering
\includegraphics[width=0.7\textwidth]{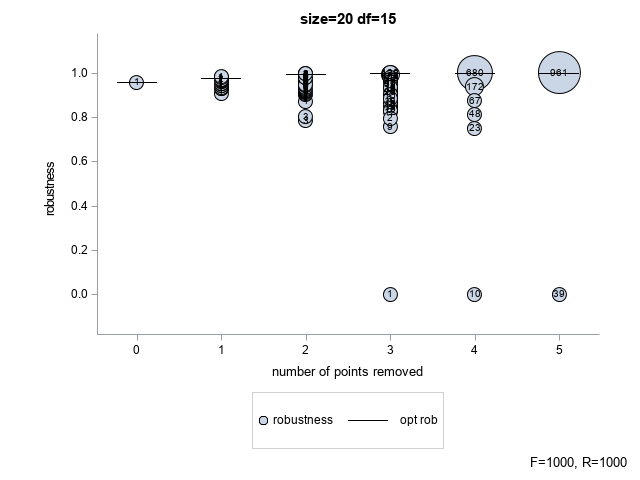}
\caption{\label{fig:bub3}Sampling distribution of robustness \emph{vs} number of points removed for the 5-factor example in Section \ref{ex:five}. For each sample size the robustness of the design selected by the proposed algorithm is represented with an horizontal line.}
\end{figure}

The third example considers five continuous variables $x_i, i=1,\ldots,5$ which take values in the interval $[-1,+1]$. The model contains the intercept, five linear terms $x_i,i=1,\ldots,5$, five quadratic terms $x_i^2,i=1,\ldots,5$ and the four interaction terms $x_1x_2, x_1x_3, x_1x_4, x_1x_5$. The number of degrees of freedom of the model is $p=1+5+5+4=15$. For this problem the (reduced) circuit basis has $276$ circuits. The robustness of a D-optimal design with with $n=20$ runs is analysed. The 20-run D-optimal design has been obtained using as candidate set the full factorial design $\{-1,+1\}^{5}$ The values of the robustness of the fractions which are obtained removing $k=1,\ldots, n-p=5$ points are computed and compared with the values of the robustness corresponding to the fractions found by the algorithm. The distributions of the robustness for each $k=1,\ldots,5$ are obtained by sampling. The results are summarized in Figure \ref{fig:bub3}. 

Table \ref{tab:id10} compares the values of the robustness of the fractions found by the algorithm ($r_*$) with the 75th, 90th and 95th percentile of the distributions of the robustness ($p_{75},p_{90},p_{95}$ respectively). It is worth noting that for each number $k$ of points removed by the initial design the algorithm provides values of robustness greater than the 95th percentile.

\begin{table}[h]
    \centering
    \begin{tabular}{c|rrrr}
    $k$ & $p_{75}$ & $p_{90}$ & $p_{95}$ & $r_*$\\
    \hline
0 & \multicolumn{4}{c}{$r_0$=0.954} \\
1 &	0.971 &	0.975 &	0.977 &	0.978 \\
2 &	0.979 &	0.989 &	0.991 &	0.994 \\
3 &	0.993 &	0.996 &	1 &	1 \\
4 &	1 &	1 &	1 &	1 \\
5 &	1 &	1 &	1 &	1 
    \end{tabular}
    \caption{Example 3: Comparison of the output of the algorithm ($r_*$) with the 75th, 90th, and 95th percentile of the distribution of the robustness ($p_{75},p_{90},p_{95}$ respectively). The value $k$ is the number of points removed by the initial design. The value corresponding to the robustness of the initial design ($r_0$) is given at $k=0$.}
    \label{tab:id10}
\end{table}

\subsection{Example 4: another design with quantitative continuous factors} \label{ex:sixteen}
\begin{figure}
\centering
\includegraphics[width=0.7\textwidth]{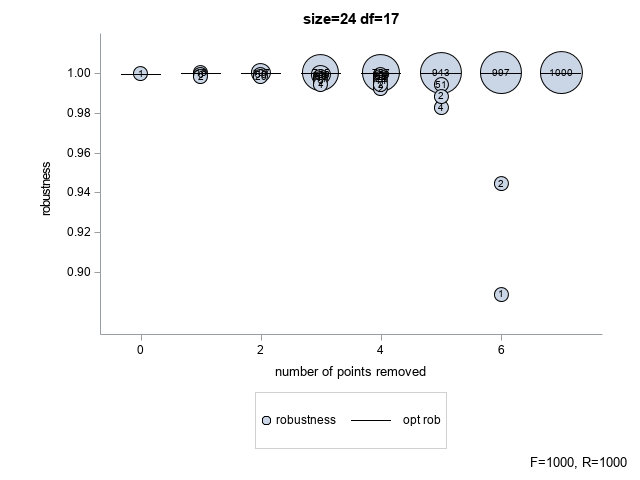}
\caption{\label{fig:bub4}Sampling distribution of robustness \emph{vs} number of points removed for the 16-factor example in Section \ref{ex:sixteen}. For each sample size the robustness of the design selected by the proposed algorithm is represented with an horizontal line.}
\end{figure}

The fourth example considers sixteen continuous variables $x_i, i=1,\ldots,16$ which take values in the interval $[-1,+1]$. The model contains the intercept and all the linear terms $x_i,i=1,\ldots,16$. The number of degrees of freedom of the model is $p=1+16=17$. For this problem the (reduced) circuit basis has $110$ circuits and, to emphasize again the relevance of the reduced circuit basis, we observe that the complete circuit basis for this problem consists of $133,883$ elements. The robustness of a D-optimal design with with $n=24$ runs is analysed. The 24-run D-optimal design has been obtained using as candidate set the full factorial design $\{-1,+1\}^{16}$. The values of the robustness of the fractions which are obtained removing $k=1,\ldots, n-p=7$ points are computed and compared with the values of the robustness corresponding to the fractions found by the algorithm. The distributions of the robustness for each $k=1,\ldots,7$ are obtained by sampling. The results are summarized in Figure \ref{fig:bub4}. 

Table \ref{tab:id11} compares the values of the robustness of the fractions found by the algorithm ($r_*$) with the 75th, 90th and 95th percentile of the distributions of the robustness ($p_{75},p_{90},p_{95}$ respectively). It is worth noting that for each number $k$ of points removed by the initial design the algorithm provides values of robustness equal to the 95th percentile (the small differences that are observed for $k=1,2$ are due to numerical approximations).

\begin{table}[h]
    \centering
    \begin{tabular}{c|rrrr}
    $k$ & $p_{75}$ & $p_{90}$ & $p_{95}$ & $r_*$\\
    \hline
0 & \multicolumn{4}{c}{$r_0$=0.9996215} \\
1 &	1 &	1 &	1 &	0.9998019 \\
2 &	1 &	1 &	1 &	0.9999241 \\
3 &	1 &	1 &	1 &	1 \\
4 &	1 &	1 &	1 &	1 \\
5 &	1 &	1 &	1 &	1 \\
6 &	1 &	1 &	1 &	1 \\
7 &	1 &	1 &	1 &	1
    \end{tabular}
    \caption{Example 4: Comparison of the output of the algorithm ($r_*$) with the 75th, 90th, and 95th percentile of the distribution of the robustness ($p_{75},p_{90},p_{95}$ respectively). The value $k$ is the number of points removed by the initial design. The value corresponding to the robustness of the initial design ($r_0$) is given at $k=0$.}
    \label{tab:id11}
\end{table}

%

\section{Final remarks} \label{sect:fin}
The main result of the paper is an algorithm for organizing the $n$ runs of a given fraction $\mathcal{F}_n$ in such a way that, if for some reasons $k$ of the $n$ runs are lost, the remaining $n-k$ runs constitutes a \emph{robust} design, $1 \leq k \leq n-p$. As shown in the simulation study, the algorithm can provide very good designs in terms of robustness for all $k \in \{1,\ldots,n-p\}$. This means that $k$ must not be defined at the design stage (it would have been extremely difficult to make an hypothesis on the number $k$ of runs that will be lost before starting the execution of the experiments). In a realistic way $k$ will take a value only after the completion of experimental activity.

The algorithm can be used with any type of initial design; the starting design $\mathcal{F}_n$ can be an orthogonal fractional factorial design or a D-optimal design or any user-defined design. 

It is worth noting that the algorithm can work even with a quite-high number of variables. The reason is that the circuits which are needed are those of the matrix $A_{\mathcal{F}_n}$ which has dimension $p \times n$ and, usually in the applications, both the number of parameters $p$ and the size of the starting fraction $\mathcal{F}_n$ are not large. 

One of the requirements is that the matrix  $A_{\mathcal{F}_n}$ must have integer values. This is always the case for models with qualitative factors. It requires some preliminary work for models with quantitative variables. With respect to this point it is worth noting that given a model matrix with real numbers is possible to build an approximate version of it with rational entries (the set of rational numbers $\mathbb{Q}$ is dense in the set of real number $\mathbb{R}$) and the approximation can be built as accurate as required. Then the model matrix can be transformed into a matrix with integer values simply by multiplying the rational matrix by a suitable integer constant. In some cases, as shown in the two final examples of Section \ref{sect:ex}, a D-optimal design used as a starting design for the algorithm contains only points with integer entries, and from the combinatorial point of view the problem reduces immediately to the qualitative case. In other scenarios, the approximation through a rational matrix is essential, but this case is outside the scope of the present paper.

We conclude with a brief account of future works in the direction of this paper. The analysis of the circuits for quantitative continuous factors and approximate algorithms for large designs where the circuit basis computation is not actually feasible have been already mentioned in the paper. Moreover, at least two important directions are suggested by the present work. From the point of view of applications, is is interesting to analyze the connections between the circuit basis and super-saturated designs, i.e., designs with a number of runs less than the number of model parameters. From a more theoretical point of view, we aim at describing the connections between robustness and circuits through the theory of matroids, which are closely related to circuits and have been studied extensively in combinatorics.

\bibliographystyle{alpha}
\bibliography{bibFRrem}

\end{document}